\documentclass[a4paper,10pt,twoside]{cpc-hepnp}
\usepackage{CJK,upgreek,fancyhdr}
\usepackage{multicol}
\usepackage{graphicx}
\usepackage{booktabs}
\usepackage{amssymb,bm,mathrsfs,bbm,amscd}
\usepackage[tbtags]{amsmath}
\usepackage{lastpage}

\begin{document}
\begin{CJK*}{GBK}{song}

\fancyhead[c]{\small Chinese Physics C~~~Vol. xx, No. x (201x) xxxxxx}
\fancyfoot[C]{\small xxxxxx-\thepage}

\footnotetext[0]{Received xx May 2017}

\title{Scalar resonant frequencies and Hawking effect of an $f(R)$ global monopole\thanks{H.S.V. is funded by the Brazilian research agencies CNPq (research Project No. 140612/2014-9) and CAPES (PDSE Process No. 88881.133092/2016-01). J.P.M.G. is funded by the CNPq through the research Project No. 150565/2016-0. V.B.B. is partially supported by the CNPq through the research Project No. 304553/2010-7}}

\author{%
      H. S. Vieira$^{1,2}$\email{Horacio.Vieira@tufts.edu}%
\quad	J. P. Morais Gra\c ca$^{2,3}$\email{jpmorais@gmail.com}%
\quad	V. B. Bezerra$^{2}$\email{valdir@fisica.ufpb.br}
}
\maketitle

\address{%
$^{1}$ Institute of Cosmology, Department of Physics and Astronomy, Tufts University, Medford, Massachusetts 02155, USA\\
$^{2}$ Departamento de F\'{i}sica, Universidade Federal da Para\'{i}ba, Caixa Postal 5008, CEP 58051-970, Jo\~{a}o Pessoa, PB, Brazil\\
$^{3}$ Instituto de F\'{i}sica, Universidade Federal do Paran\'{a}, Curitiba, PR, Brazil\\
}

\begin{abstract}
Massive scalar fields are considered in the gravitational field produced by a Schwarzschild black hole with a global monopole in $f(R)$ gravity. The exact solution of the radial part of the Klein-Gordon equation in this background is obtained and is given in terms of the general Heun functions. We apply the properties of the general Heun functions to study the Hawking radiation and the resonant frequencies of scalar particles.
\end{abstract}



\footnotetext[0]{\hspace*{-3mm}\raisebox{0.3ex}{$\scriptstyle\copyright$}2013
Chinese Physical Society and the Institute of High Energy Physics
of the Chinese Academy of Sciences and the Institute
of Modern Physics of the Chinese Academy of Sciences and IOP Publishing Ltd}%


%
%
\section{Introduction}
In principle, a way to get much information about the physics of black holes, and other similar spacetimes, is to analyze their interaction with quantum fields, especially scalar fields \cite{PhysLettB.751.34,AnnPhys.362.363,AnnPhys.362.576,EurPhysJPlus.131.427,ClassQuantumGrav.33.225011,ModPhysLettA.31.1650057,AdvHighEnergyPhys.2017.8934691}. One such source of information is the Resonant Frequencies (RFs) emitted by a black hole via its interaction with different fields \cite{AnnPhys.373.28,PhysRevD.94.084040}.

Since the expansion of the universe was discovered, several new theories of gravity have been developed to take  this fact into account. One of these is $f(R)$ gravity, in which a function of the Ricci scalar, $R$, is introduced in the Einstein-Hilbert action \cite{RevModPhys.82.451}. Other studies concerning alternative theories of gravity as well as their consequences can be found in Refs.~\cite{PhysLettB.91.99,PhysLettB.681.74,ModPhysLettA.30.1550156,ClassQuantumGrav.33.055004}.

In the 21st century, the most important special functions are the Heun functions. Indeed, there are a large number of applications of these functions in different areas of physics \cite{arXiv:1101.0471}, in particular problems concerning the interaction between scalar fields and gravitational backgrounds \cite{ClassQuantumGrav.31.045003,AnnPhys.528.264}. Otherwise, without the use of these functions, scalar solutions would be possible only for asymptotic regimes.

In the present work, we apply the general Heun equation to obtain the solution of the radial Klein-Gordon equation for a massive scalar field in a Schwarzschild black hole with a global monopole in $f(R)$ gravity. This solution, which is given in terms of the general Heun functions, is used to examine the resonant frequencies and the Hawking radiation of scalar waves.

The richness of the new ideas and the perspective of describing the open problems in cosmology justify the interest in the study of $f(R)$ gravity. From a cosmological point of view, the $f(R)$ theory makes it possible, in principle, to explain both the late time cosmic speed-up and the early time inflationary scenario in one model, without the introduction of an ad hoc cosmological constant \cite{ModPhysLettA.30.1550217}. Furthermore, $f(R)$ gravity can be seen as an effective theory that introduces corrections to general relativity, such as, for example, to explain the rotation of galaxies \cite{RussPhysJ.48.940}, and the Starobinsky model that can be used to fit some results on Cosmic Microwave Background data \cite{AstronAstrophys.571.A22}.

On the other hand, motivated by the idea of constructing a theory combining quantum physics and general relativity, the $f(R)$ theory can be used to study the interaction between scalar particles and one of their black hole solutions, namely, the Schwarzschild black hole with a global monopole \cite{GenRelativGravit.48.38}. This can lead us to find the influence of modified gravity on the thermal properties of black holes \cite{PhysRevD.87.044002}, for example.

This paper is organized as follows. In Section 2, we present the background under consideration. In Section 3, we obtain the solution of the Klein-Gordon equation for a massive scalar field in this spacetime. In Section 4, we discuss the Hawking radiation effect. In Section 5, we obtain the resonant frequencies. Finally, in Section 6, the conclusions are given.
%
%
\section{Schwarzschild black hole with a global monopole in $f(R)$ gravity}
The metric generated by a static and spherically symmetric spacetime in $f(R)$ gravity corresponds to the Schwarzschild black hole with a global monopole \cite{GenRelativGravit.48.38}, whose line element is given by

\begin{equation}
ds^{2}=\Delta\ dt^{2}-\Delta^{-1}\ dr^{2}-r^{2}(d\theta^{2}+\sin^2\theta\ d\phi^{2})\ ,
\label{eq:metric_spherically_monopole}
\end{equation}
with
\begin{equation}
\Delta=1-\frac{2M}{r}-8\pi\eta^{2}-\psi_{0}r\ ,
\label{eq:Delta_metric_spherically_monopole}
\end{equation}
where $M$ is the total mass of the Schwarzschild black hole with a global monopole. The deviation from general relativity is indicated by the constant $\psi_{0}$, which is assumed to have values such that $|\psi_{0}r| \ll 1$ in the weak field approximation constrained equation $\frac{df(R)}{dR}=1+\psi_{0}r$. The parameter $\eta$ is related to the scalar field vacuum expectation value and we will assume that $\eta^{2} \ll 1$ (for a review, see Ref.~\cite{IntJModPhysConfSer.03.446} and references therein).

The horizon surface equation is obtained from Eq.~(\ref{eq:Delta_metric_spherically_monopole}), namely, under the condition
\begin{equation}
\Delta=(r-r_{+})(r-r_{-})=0\ .
\label{eq:surface_hor_spherically_monopole}
\end{equation}
The solutions of Eq.~(\ref{eq:surface_hor_spherically_monopole}) are
\begin{equation}
r_{\pm}=\frac{1-8 \pi  \eta ^2\pm\sqrt{(1-8 \pi  \eta ^2)^2-8 M \psi_{0} }}{2 \psi_{0} }\ ,
\label{eq:sol_surface_hor_spherically_monopole_1}
\end{equation}
and correspond to the event horizons of the background under consideration; $r_{-}$ is the interior event horizon and $r_{+}$ is the cosmological event horizon.
%
%
\section{Exact solution of the Klein-Gordon equation}
Now, let us consider the covariant Klein-Gordon equation for a massive scalar field in a curved spacetime. In this case, we can write the Klein-Gordon equation as
\begin{equation}
\biggl[\frac{1}{\sqrt{-g}}\partial_{\sigma}(g^{\sigma\tau}\sqrt{-g}\partial_{\tau})+\mu_{0}^{2}\biggr]\Psi=0\ ,
\label{eq:Klein-Gordon}
\end{equation}
where $\mu_{0}$ is the mass of the scalar particle. Note that the units $G \equiv c \equiv \hbar \equiv k_{B} \equiv 1$ were chosen.

Substituting Eq.~(\ref{eq:metric_spherically_monopole}) into Eq.~(\ref{eq:Klein-Gordon}), we obtain
\begin{equation}
\biggl[-\frac{r^{2}}{\Delta}\frac{\partial^{2}}{\partial t^{2}}+\frac{\partial}{\partial r}\biggl(r^{2}\Delta\frac{\partial}{\partial r}\biggr)-\mathbf{L}^{2}-\mu_{0}^{2}r^{2}\biggr]\Psi=0\ ,
\label{eq:mov_spherically_monopole}
\end{equation}
where $\mathbf{L}^{2}$ is the angular momentum operator given by
\begin{equation}
\mathbf{L}^{2}=-\frac{1}{\sin\theta}\frac{\partial}{\partial \theta}\biggl(\sin\theta\frac{\partial}{\partial \theta}\biggr)-\frac{1}{\sin^{2}\theta}\frac{\partial^{2}}{\partial\theta^{2}}\ .
\label{eq:momentum_operator}
\end{equation}
Due to the time independence and symmetry of the spacetime under consideration with respect to rotation, the solution of Eq.~(\ref{eq:mov_spherically_monopole}) can be written as
\begin{equation}
\Psi=\Psi(\mathbf{r},t)=R(r)Y_{l}^{m}(\theta,\phi)\mbox{e}^{-i\omega t}\ ,
\label{eq:ansatz_solution_spherically_monopole}
\end{equation}
where $Y_{l}^{m}(\theta,\phi)$ are called spherical harmonics, with $l=\{0,1,2,...\}$ being the orbital quantum number and $|m| \leq l$ is the azimuthal quantum number. The energy (frequency) is taken as $\omega > 0$, which corresponds to the flux of particles at infinity.

Substituting Eq.~(\ref{eq:ansatz_solution_spherically_monopole}) into Eq.~(\ref{eq:mov_spherically_monopole}), we find that \cite{GenRelativGravit.48.38}
\begin{equation}
\frac{d}{dr}\biggl(r^{2}\Delta\frac{dR}{dr}\biggr)+\biggl(\frac{r^{2}\omega^{2}}{\Delta}-\lambda_{lm}-\mu_{0}^{2}r^{2}\biggr)R=0\ ,
\label{eq:mov_radial_spherically_monopole}
\end{equation}
where $\lambda_{lm}=l(l+1)$ is a constant.
%
%
\subsection{Radial equation}
Now, let us obtain the exact and general solution for the radial part of the Klein-Gordon equation given by Eq.~(\ref{eq:mov_radial_spherically_monopole}).

To solve the radial part of the Klein-Gordon equation, we use Eq.~(\ref{eq:surface_hor_spherically_monopole}) and write down Eq.~(\ref{eq:mov_radial_spherically_monopole}) as


\begin{equation}
\frac{d^{2}R}{dr^{2}}+\biggl(\frac{2}{r}+\frac{1}{r-r_{+}}+\frac{1}{r-r_{-}}\biggr)\frac{dR}{dr}+\biggr\{\frac{1}{r (r-r_{+}) (r-r_{-})}\biggr[-\mu_{0} ^2 r-\frac{\lambda_{lm} }{r}+\frac{r_{+} \omega ^2}{(r_{+}-r_{-})}\frac{1}{r-r_{+}}-\frac{r_{-} \omega ^2}{(r_{+}-r_{-})}\frac{1}{r-r_{-}}\biggr]\biggr\}R=0\ .
\label{eq:mov_radial_spherically_monopole_r}
\end{equation}


This equation has singularities at $r=(a_{1},a_{2},a_{3},a_{\infty})$ $=(r_{-},r_{+},0,\infty)$. The transformation of Eq.~(\ref{eq:mov_radial_spherically_monopole_r}) to a Heun-type equation is achieved by setting
\begin{equation}
x=\frac{r-a_{1}}{a_{2}-a_{1}}=\frac{r-r_{-}}{r_{+}-r_{-}}\ ,
\label{eq:homog_subs_radial_spherically_monopole_x}
\end{equation}
which transforms $(a_{1},a_{2}) \mapsto (0,1)$, and the remaining singularity is transformed to $x=a$, where
\begin{equation}
a \equiv \frac{a_{3}-a_{1}}{a_{2}-a_{1}}=\frac{-r_{-}}{r_{+}-r_{-}}\ .
\label{eq:singularity_a_spherically_monopole}
\end{equation}
This is an homographic substitution which has the following asymptotic regimes: $x \rightarrow 0 \Rightarrow r \rightarrow r_{-}$ and $x \rightarrow \infty \Rightarrow r \rightarrow \infty$. Then, it is easy to see that this transformation covers the entire spacetime region of the Schwarzschild black hole with a global monopole in the $f(R)$ theory of gravity.

Thus, we can write Eq.~(\ref{eq:mov_radial_spherically_monopole_r}) as


\begin{eqnarray}
&& \frac{d^{2}R}{dx^{2}}+\biggl(\frac{1}{x}+\frac{1}{x-1}+\frac{2}{x-a}\biggr)\frac{dR}{dx}+\biggl\{\frac{2 a^2 \omega ^2 +\mu_{0} ^2 r_{-}^2+\lambda_{lm}}{r_{-}^2 }\frac{1}{x}+\frac{-(2 a^2 \omega ^2+\mu_{0} ^2 r_{-}^2)(a-1)^2-a^2 \lambda_{lm}}{(a-1)^2 r_{-}^2}\frac{1}{x-1}\nonumber\\
&& +\frac{(2 a-1) \lambda_{lm} }{(a-1)^2 r_{-}^2 }\frac{1}{x-a}+\frac{a^2 \omega ^2}{r_{-}^2 }\frac{1}{x^2}+\frac{a^2 \omega ^2}{r_{-}^2 }\frac{1}{(x-1)^2}-\frac{a \lambda_{lm} }{(a-1) r_{-}^2 }\frac{1}{(x-a)^2}\biggr\}R=0\ .
\label{eq:mov_radial_spherically_monopole_x}
\end{eqnarray}


Now, let us perform a transformation in order to reduce the powers of the terms proportional to $1/x^{2}$, $1/(x-1)^{2}$, and $1/(x-a)^{2}$. This transformation is a F-homotopic transformation of the dependent variable, $R(x) \mapsto U(x)$, such that
\begin{equation}
R(x)=x^{A_{1}}(x-1)^{A_{2}}(x-a)^{A_{3}}U(x)\ ,
\label{eq:F-homotopic_mov_radial_spherically_monopole_x}
\end{equation}
where the coefficients $A_{1}$, $A_{2}$, and $A_{3}$ are given by
\begin{equation}
A_{1}=\frac{i a \omega }{r_{-}}\ ,
\label{eq:A1_radial_spherically_monopole_x}
\end{equation}
\begin{equation}
A_{2}=\frac{i a \omega }{r_{-}}\ ,
\label{eq:A2_radial_spherically_monopole_x}
\end{equation}
\begin{equation}
A_{3}=\frac{\{(a-1)r_{-}^{2}[4 a \lambda_{lm} +(a-1)r_{-}^2]\}^{1/2}-r_{-}^2(a-1)}{2(a-1)r_{-}^2}\ ,
\label{eq:A3_radial_spherically_monopole_x}
\end{equation}
The function $U(x)$ satisfies the following equation
\begin{eqnarray}
&& \frac{d^{2}U}{dx^{2}}+\biggl[\frac{2 A_{1}+1}{x}+\frac{2 A_{2}+1}{x-1}+\frac{2 (A_{3}+1)}{x-a}\biggr]\frac{dU}{dx}\nonumber\\
&& +\frac{A_{4}x-A_{5}}{x(x-1)(x-a)}U=0\ ,
\label{eq:mov_radial_spherically_monopole_x_U}
\end{eqnarray}
where the coefficients $A_{4}$ and $A_{5}$ are given by
\begin{eqnarray}
A_{4} & = & \frac{(a-1)[r_{-}^2 A_{2} (2 A_{3}+3)+2 r_{-}^2 A_{3} - \mu_{0} ^2 r_{-}^2]+a \lambda_{lm}}{(a-1) r_{-}^2}\nonumber\\
 & + & \frac{(a-1)[r_{-}^2 A_{1} (2 A_{2}+2 A_{3}+3)-2 a^2 \omega ^2]}{(a-1) r_{-}^2}\ ,
\label{eq:A4_radial_spherically_monopole_x}
\end{eqnarray}
\begin{eqnarray}
A_{5} & = & A_{3}+A_{1} (2 a A_{2}+a+2 A_{3}+2)\nonumber\\
 & + & a \biggl(A_{2}-\mu_{0} ^2-\frac{\lambda_{lm} }{r_{-}^2}\biggr)-\frac{2 a^3 \omega ^2}{r_{-}^2}\ .
\label{eq:A5_radial_spherically_monopole_x}
\end{eqnarray}

Equation (\ref{eq:mov_radial_spherically_monopole_x_U}) is similar to the general Heun equation \cite{Ronveaux:1995}, namely,
\begin{equation}
\frac{d^{2}U}{dx^{2}}+\biggl(\frac{\gamma}{x}+\frac{\delta}{x-1}+\frac{\epsilon}{x-a}\biggr)\frac{dU}{dx}+\frac{\alpha\beta x-q}{x(x-1)(x-a)}U=0\ ,
\label{eq:canonical_form_general_Heun}
\end{equation}
where $U(x)=\mbox{HeunG}(a,q;\alpha,\beta,\gamma,\delta;x)$ are the general Heun functions. This is a Fuchsian type equation with regular singularities at $x=(0,1,a,\infty)$. The general Heun function is simultaneously a local Frobenius solution around a singularity $x=a_{i}$ and a local Frobenius solution around $x=a_{j}$, so that it is analytic in some domain including both these singularities. The parameters $\alpha$, $\beta$, $\gamma$, $\delta$, $\epsilon$, $q$, $a$ are generally complex, arbitrary (except that $a \neq 0,1$), and related by
\begin{equation}
\gamma+\delta+\epsilon=\alpha+\beta+1\ .
\label{eq:parameters_relation_general_Heun}
\end{equation}
If $\gamma \neq 0,-1,-2,...$, then from the Fuchs-Frobenius Theory, it follows that $\mbox{HeunG}(a,q;\alpha,\beta,\gamma,\delta;x)$ exists, is analytic in the disk $|x| < 1$, corresponds to exponent $0$ at $x=0$ and assumes the value $1$ there, and has the Maclaurin expansion
\begin{equation}
\mbox{HeunG}(a,q;\alpha,\beta,\gamma,\delta;x)=\sum_{j=0}^{\infty}b_{j}x^{j}, \quad |x| < 1\ ,
\label{eq:serie_HeunG_todo_x}
\end{equation}
where $b_{0}=1$, and
\begin{eqnarray}
	a\gamma b_{1}-qb_{0}=0\ ,\nonumber\\
	X_{j}b_{j+1}-(Q_{j}+q)b_{j}+P_{j}b_{j-1}=0, \quad j \geq 1\ ,
\label{eq:recursion_General_Heun}
\end{eqnarray}
with
\begin{eqnarray}
	P_{j}=(j-1+\alpha)(j-1+\beta)\ ,\nonumber\\
	Q_{j}=j[(j-1+\gamma)(1+a)+a\delta+\epsilon]\ ,\nonumber\\
	X_{j}=a(j+1)(j+\gamma)\ .
\label{eq:P_Q_X_recursion_General_Heun}
\end{eqnarray}

Thus, the general solution of the radial part of the Klein-Gordon equation for a massive scalar particle in the Schwarzschild black hole with a global monopole spacetime in $f(R)$ gravity, given by Eq.~(\ref{eq:mov_radial_spherically_monopole_x}), valid in the exterior region of the event horizon $r_{-}$, can be written as
\begin{eqnarray}
R(x) & = & x^{\frac{1}{2}(\gamma-1)}(x-1)^{\frac{1}{2}(\delta-1)}(x-a)^{\frac{1}{2}(\epsilon-2)}\nonumber\\
 & \times & \{C_{1}\ \mbox{HeunG}(a,q;\alpha,\beta,\gamma,\delta;x)\nonumber\\
 & + & C_{2}\ x^{1-\gamma}\ \mbox{HeunG}(a,q_{1};\alpha_{1},\beta_{1},\gamma_{1},\delta;x)\}\ ,\nonumber\\
\label{eq:general_solution_radial_spherically_monopole_x}
\end{eqnarray}
where $C_{1}$ and $C_{2}$ are constants, and the parameters $\alpha$, $\beta$, $\gamma$, $\delta$, $\epsilon$, and $q$ are now given by
\begin{eqnarray}
\alpha & = & \frac{[4 a \lambda_{lm}(a-1) +(a-1)^{2} r_{-}^2]^{1/2}}{2 (a-1) r_{-}}\nonumber\\
 & + & \frac{[(4 \mu_{0} ^2+9) (a-1)^{2} r_{-}^2]^{1/2}}{2 (a-1) r_{-}}\nonumber\\
 & + & \frac{4 i a \omega(a-1) +2 (a-1) r_{-}}{2 (a-1) r_{-}}\ ,
\label{eq:alpha_radial_spherically_monopole_x}
\end{eqnarray}
\begin{eqnarray}
\beta & = & \frac{[4 a \lambda_{lm}(a-1) +(a-1)^{2} r_{-}^2]^{1/2}}{2 (a-1) r_{-}}\nonumber\\
 & - & \frac{[(4 \mu_{0} ^2+9) (a-1)^{2} r_{-}^2]^{1/2}}{2 (a-1) r_{-}}\nonumber\\
 & + & \frac{4 i a \omega(a-1) +2 (a-1) r_{-}}{2 (a-1) r_{-}}\ ,
\label{eq:beta_radial_spherically_monopole_x}
\end{eqnarray}
\begin{equation}
\gamma=1+\frac{2 i a \omega }{r_{-}}\ ,
\label{eq:gamma_radial_spherically_monopole_x}
\end{equation}
\begin{equation}
\delta=1+\frac{2 i a \omega }{r_{-}}\ ,
\label{eq:delta_radial_spherically_monopole_x}
\end{equation}
\begin{equation}
\epsilon=\frac{[4 a \lambda_{lm}(a-1) +(a-1)^{2} r_{-}^2]^{1/2}+(a-1) r_{-}}{(a-1) r_{-}}\ ,
\label{eq:eta_radial_spherically_monopole_x}
\end{equation}
\begin{eqnarray}
q & = & \frac{r_{-} \{4 i a^3 \omega -2 i a \omega(a+1)}{2 (a-1) r_{-}^2}\nonumber\\
 & + & \frac{[4 a \lambda_{lm}(a-1) +(a-1)^{2} r_{-}^2]^{1/2} \}}{2 (a-1) r_{-}^2}\nonumber\\
 & + & \frac{r_{-}^2 (1-a)(2 a \mu_{0} ^2-1)}{2 (a-1) r_{-}^2}\nonumber\\
 & + & \frac{2 a \{\omega  \{-4 a^2 \omega(a-1)}{2 (a-1) r_{-}^2}\nonumber\\
 & + & \frac{i [4 a \lambda_{lm}(a-1) +(a-1)^{2} r_{-}^2]^{1/2}\}}{2 (a-1) r_{-}^2}\nonumber\\
 & + & \frac{\lambda_{lm}(1-a)\}}{2 (a-1) r_{-}^2}\ .
\label{eq:q_radial_spherically_monopole_x}
\end{eqnarray}
The parameters $\alpha_{1}$, $\beta_{1}$, $\gamma_{1}$, and $q_{1}$ are given by
\begin{equation}
\alpha_{1}=\alpha+1-\gamma\ ,
\label{eq:alpha_1_general_Heun}
\end{equation}
\begin{equation}
\beta_{1}=\beta+1-\gamma\ ,
\label{eq:beta_1_general_Heun}
\end{equation}
\begin{equation}
\gamma_{1}=2-\gamma\ .
\label{eq:gamma_1_general_Heun}
\end{equation}
\begin{equation}
q_{1}=q+(\alpha\delta+\epsilon)(1-\gamma)\ .
\label{eq:q_1_general_Heun}
\end{equation}

At this point, we can compare the result obtained  with those given by Refs.~\cite{ClassQuantumGrav.31.045003,AnnPhys.350.14}. First, it is worth paying attention to the difference between the functional form of the general solution of the radial equation obtained analytically, given by Eq.~(\ref{eq:general_solution_radial_spherically_monopole_x}) in terms of the general Heun functions, and the solutions obtained, also analytically, in \cite{ClassQuantumGrav.31.045003,AnnPhys.350.14}, given by their Eqs.~(31) and (78) (both in the case $a=Q=e=0$), respectively, in terms of the confluent Heun functions. It is always possible to reduce the general Heun function to the confluent Heun function by a general process of confluence but the limiting process generally alters the solutions qualitatively. Then, the results in this paper can recover just the functional form of the results presented in Refs. \cite{ClassQuantumGrav.31.045003,AnnPhys.350.14}.

In what follows, we will use this analytical solution for the radial part of the Klein-Gordon equation and the presented properties of the general Heun function to study the Hawking effect and the resonant frequencies.
%
%
\section{Hawking radiation}
We will consider the massive scalar field near the interior event horizon  in order to discuss the Hawking radiation.

From Eqs.~(\ref{eq:homog_subs_radial_spherically_monopole_x}) and (\ref{eq:serie_HeunG_todo_x}) we can see that the radial solution given by Eq.~(\ref{eq:general_solution_radial_spherically_monopole_x}), near the interior event horizon, that is, when $r \rightarrow r_{-}$ (which implies that $x \rightarrow 0$), behaves asymptotically as
\begin{equation}
R(r) \sim C_{1}\ (r-r_{-})^{\frac{1}{2}(\gamma-1)}+C_{2}\ (r-r_{-})^{-\frac{1}{2}(\gamma-1)}\ ,
\label{eq:exp_0_solucao_geral_radial_spherically_monopole_x}
\end{equation}
where we are considering contributions only of the first term in the expansion, and all constants are included in $C_{1}$ and $C_{2}$. Thus, considering the time factor, near the black hole event horizon $r_{-}$, this solution is given by
\begin{equation}
\Psi=\mbox{e}^{-i \omega t}(r-r_{-})^{\pm\frac{1}{2}(\gamma-1)}\ .
\label{eq:sol_onda_radial_spherically_monopole_x}
\end{equation}
From Eq.~(\ref{eq:gamma_radial_spherically_monopole_x}), we obtain
\begin{equation}
\frac{1}{2}(\gamma-1)=\frac{i}{2\kappa_{-}}\omega\ ,
\label{eq:expoente_rad_Hawking_spherically_monopole_x}
\end{equation}
where $\kappa_{-}$ is the surface gravity of the black hole (or the gravitational acceleration on the background horizon surface $r_{-}$) given by
\begin{equation}
\kappa_{-} \equiv \frac{1}{2r_{-}^{2}}\left.\frac{d\Delta}{dr}\right|_{r=r_{-}}=\frac{r_{-}-r_{+}}{2r_{-}^{2}}\ .
\label{eq:acel_grav_ext_spherically_monopole}
\end{equation}

Therefore, on the black hole interior horizon surface, the ingoing and outgoing wave solutions are
\begin{equation}
\Psi_{in}=\mbox{e}^{-i \omega t}(r-r_{-})^{-\frac{i}{2\kappa_{-}}\omega}\ ,
\label{eq:sol_in_1_spherically_monopole_x}
\end{equation}
\begin{equation}
\Psi_{out}(r>r_{-})=\mbox{e}^{-i \omega t}(r-r_{-})^{\frac{i}{2\kappa_{-}}\omega}\ .
\label{eq:sol_out_2_spherically_monopole_x}
\end{equation}
Using the definition of the tortoise and advanced Eddington-Finkelstein coordinates
\begin{equation}
r_{*}=\frac{1}{2\kappa_{-}}\ln(r-r_{-})\ ,
\label{eq:coord_tortoise}
\end{equation}
\begin{equation}
v=t+r_{*}\ ,
\label{eq:coord_Eddington-Finkelstein}
\end{equation}
we can rewrite $\Psi_{out}$, near the interior event horizon $r_{-}$, as
\begin{equation}
\Psi_{out}(r>r_{-})=\mbox{e}^{-i \omega v}(r-r_{-})^{\frac{i}{\kappa_{-}}\omega}\ .
\label{eq:sol_out_2_spherically_monopole_x_tortoise}
\end{equation}

This solution is not analytical in the exterior event horizon $r=r_{-}$. According to the Damour-Ruffini method \cite{PhysRevD.14.332}, by analytic continuation, rotating $-\pi$ through the lower-half complex $r$ plane, namely,
\begin{equation}
(r-r_{-}) \rightarrow \left|r-r_{-}\right|\mbox{e}^{-i\pi}=(r_{-}-r)\mbox{e}^{-i\pi}\ ,
\label{eq:rel_3_spherically_monopole_x}
\end{equation}
we can extend $\Psi_{out}$ from the outside of the black hole into the inside of the black hole, and get the outgoing decay rate (or the relative scattering probability) of the scalar wave at the interior event horizon surface $r=r_{-}$
\begin{equation}
\Gamma_{-}=\left|\frac{\Psi_{out}(r>r_{-})}{\Psi_{out}(r<r_{-})}\right|^{2}=\mbox{e}^{-\frac{2\pi}{\kappa_{-}}\omega}\ ,
\label{eq:taxa_refl_spherically_monopole_x}
\end{equation}
This is the relative probability of creating a particle-antiparticle pair just outside the interior horizon.

According to the Sannan heuristic derivation \cite{GenRelativGravit.20.239}, the mean number of particles emitted in a given mode, $\bar{N}_{\omega}$, can be obtained from the expression for the relative scattering probability, $\Gamma_{-}$, as follows:
\begin{equation}
\bar{N}_{\omega}=\frac{\Gamma_{-}}{1-\Gamma_{-}}=\frac{1}{\mbox{e}^{\frac{\hbar\omega}{k_{B}T_{-}}}-1}\ ,
\label{eq:spectrum_rad_spherically_monopole_x}
\end{equation}
where
\begin{equation}
T_{-}=\frac{\kappa_{-}}{2\pi}=\frac{r_{-}-r_{+}}{4\pi r_{-}^{2}}
\label{eq:temp_Hawking_spherically_monopole_x}
\end{equation}
is the Hawking radiation temperature in geometric units. Equation (\ref{eq:spectrum_rad_spherically_monopole_x}) is exactly the resulting Hawking radiation spectrum for scalar particles being radiated from a black hole with a global monopole in $f(R)$ gravity, where Boltzmann's and Planck's constants are reintroduced.

Therefore, we can see that the resulting Hawking radiation spectrum of scalar particles has a thermal character, analogous to the black body spectrum, where $k_{B}T_{-}=\hbar\kappa_{-}/2\pi$.

By integrating the above spectrum (or distribution function) over all $\omega$'s, we can obtain the Hawking flux for massive scalar particles. It is given by
\begin{equation}
\mbox{Flux}=\frac{1}{2\pi}\int_{0}^{\infty}{\bar{N}_{\omega}\ \omega\ d\omega}=\frac{1}{2\pi}\int_{0}^{\infty}{\frac{\omega\ d\omega}{\mbox{e}^{\frac{2\pi}{\kappa_{-}}\omega}-1}}=\frac{\kappa_{-}^{2}}{48\pi}\ .
\label{eq:flux_rad_spherically_monopole_x}
\end{equation}

Once again, we can compare the results obtained  with those of Ref.~\cite{AnnPhys.350.14}. The surface gravity of the black hole, $\kappa_{-}$, and the Hawking radiation temperature, $T_{-}$, given by Eqs.~(\ref{eq:acel_grav_ext_spherically_monopole}) and (\ref{eq:temp_Hawking_spherically_monopole_x}), respectively, have magnitudes bigger than those obtained in Ref.~\cite{AnnPhys.350.14}, given by its Eqs.~(10) and (110). Then, the mean number of particles emitted in a given mode, $\bar{N}_{\omega}$, for scalar particles being radiated from a Schwarzschild black hole with a global monopole in $f(R)$ gravity is bigger than that from a Schwarzschild black hole in general relativity. Furthermore, these two quantities are negative in the background under consideration.
%
%
\section{Resonant frequencies}
In this section, we follow the recently developed technique of Ref.~\cite{AnnPhys.373.28} for computing the RFs for massive scalar waves propagating in a Schwarzschild black hole with a global monopole in $f(R)$ gravity.

The RFs are associated with the solution given by Eq.~(\ref{eq:general_solution_radial_spherically_monopole_x}) under certain boundary conditions, that is, the radial solution should be finite on the interior horizon and well behaved at asymptotic infinity. The latter condition requires that $R(x)$ must have a polynomial form. Indeed, the function $\mbox{HeunG}(a,q;\alpha,\beta,\gamma,\delta;x)$ becomes a polynomial of degree $n$ if
\begin{equation}
\alpha=-n\ ,
\label{eq:condiction_poly_General_Heun}
\end{equation}
with $n=0,1,2,\ldots$.

Using Eq.~(\ref{eq:alpha_radial_spherically_monopole_x}), we therefore find that the RFs for a massive scalar particle in the background under consideration are given by
\begin{equation}
\omega_{n}=i\biggl\{\frac{[4 a \lambda_{lm}(a-1) +(a-1)^{2} r_{-}^2]^{1/2}+[(a-1)^{2} (4 \mu_{0} ^2+9) r_{-}^2]^{1/2}}{4 (a-1) a}+\frac{2 (a-1) (n+1) r_{-}}{4 (a-1) a}\biggr\}\ ,
\label{eq:Energy_levels_spherically_monopole_x}
\end{equation}
where the quantum number $n$ is a positive integer or zero.

This is a nontrivial quantization law, because it gives a complex number. We remark that the eigenvalues given by Eq.~(\ref{eq:Energy_levels_spherically_monopole_x}) are degenerate, since that there is a dependence on the eigenvalue $\lambda_{lm}$.

The resonant frequencies for $n=\{0,1,2,3\}$, $0 \leq l \leq n$, $\psi_{0}=0.02$, $M=1$ and $\mu_{0}=\{0,0.1,0.2,0.3\}$ are shown in Tables \ref{tab:Tab1_monopole} and \ref{tab:Tab2_monopole} for two different values of the parameter $\eta$, namely, $8\pi\eta^{2}=10^{-5}$ and $8\pi\eta^{2}=0.02$, respectively. These are the same set of values that were used in our recent work \cite{GenRelativGravit.48.38} for the quasinormal modes.


\begin{center}
\tabcaption{ \label{tab:Tab1_monopole} Values of the resonant frequencies for $l=0$, $\psi_{0}=0.02$, and $8\pi\eta^{2}=10^{-5}$ with different values of mass $\mu_{0}$ and principal quantum number $n$. The units are in multiples of the total mass $M$.}
\normalsize
\begin{tabular*}{170mm}{@{\extracolsep{\fill}}lllll@{\extracolsep{\fill}}}
\toprule
	$a=-0.04554$	& \ $\mu_{0}=0.0$				& \ $\mu_{0}=0.1$				& \ $\mu_{0}=0.2$				& \ $\mu_{0}=0.3$				\\
	$n$						&	\ \ $\Im(\omega_{n})$	&	\ \ $\Im(\omega_{n})$	&	\ \ $\Im(\omega_{n})$	&	\ \ $\Im(\omega_{n})$	\\\hline
	0							&	\ 22.9126							& \ 22.9888							& \ 23.2167							&	\ 23.5932							\\
	1							&	1.049e-8							& \ 0.07629							&	\ 0.30415							&	\ 0.68063							\\
	2							&	-22.9126							& -22.8363							&	-22.6084							&	-22.2319							\\
	3							&	-45.8252							& -45.7489							&	-45.5210							&	-45.1445							\\
\bottomrule
\end{tabular*}
\end{center}
\begin{center}
\tabcaption{ \label{tab:Tab2_monopole} Values of the resonant frequencies for $l=0$, $\psi_{0}=0.02$, and $8\pi\eta^{2}=0.02$ with different values of mass $\mu_{0}$ and principal quantum number $n$. The units are in multiples of the total mass $M$.}
\normalsize
\begin{tabular*}{170mm}{@{\extracolsep{\fill}}lllll@{\extracolsep{\fill}}}
\toprule
	$a=-0.04769$	& \ $\mu_{0}=0.0$				& \ $\mu_{0}=0.1$				& \ $\mu_{0}=0.2$				& \ $\mu_{0}=0.3$				\\
	$n$						&	\ \ $\Im(\omega_{n})$	&	\ \ $\Im(\omega_{n})$	&	\ \ $\Im(\omega_{n})$	&	\ \ $\Im(\omega_{n})$	\\\hline
	0							&	\ 22.3662							& \ 22.4407							& \ 22.6631							&	\ 23.0306							\\
	1							&	2.501e-8							& \ 0.07447							&	\ 0.29690							&	\ 0.66440							\\
	2							&	-22.3662							& -22.2917							&	-22.0693							&	-21.7018							\\
	3							&	-44.7325							& -44.6580							&	-44.4356							&	-44.0681							\\
\bottomrule
\end{tabular*}
\end{center}


Once again, we can compare the results obtained  with those of Ref.~\cite{AnnPhys.373.28}. The RFs, $\omega_{n}$, given by Eq.~(\ref{eq:Energy_levels_spherically_monopole_x}), decrease with quantum number $n$ for massless scalar fields ($\mu_{0}=0.0$). On the other hand, the RFs from a Schwarzschild black hole in general relativity, given by Eq.~(61) in Ref.~\cite{AnnPhys.373.28}, increase with quantum number $n$. Furthermore, this quantity becomes negative in the background under consideration.

In Figs.~(\ref{fig:Fig1_monopole}), (\ref{fig:Fig2_monopole}), and (\ref{fig:Fig3_monopole}) we present the resonant frequencies as functions of $l$, $\mu_{0}$, and $n$, respectively, for $8\pi\eta^{2}=10^{-5}$, as well as in Figs.~(\ref{fig:Fig4_monopole}), (\ref{fig:Fig5_monopole}), and (\ref{fig:Fig6_monopole}) for $8\pi\eta^{2}=0.02$. This is the same set of variables that was used in Refs.~\cite{PhysRevD.73.124040,CentEurJPhys.6.194} for the quasinormal modes.

\begin{center}
		\includegraphics[scale=0.40]{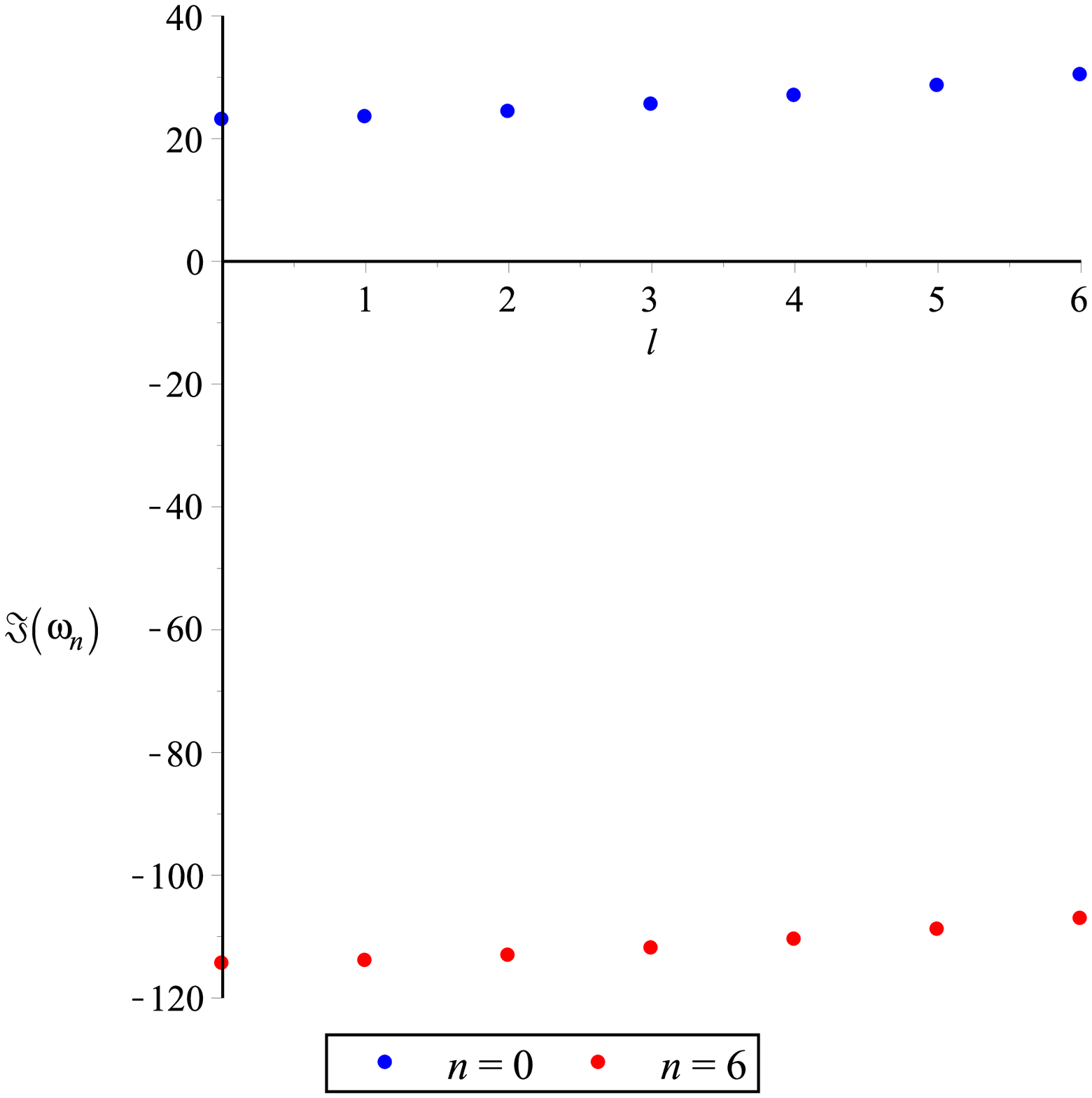}
	\figcaption{Scalar resonant frequencies for $\mu_{0}=0.1$, $\psi_{0}=0.02$, and $8\pi\eta^{2}=10^{-5}$. The units are in multiples of the total mass $M$.}
	\label{fig:Fig1_monopole}
%
		\includegraphics[scale=0.40]{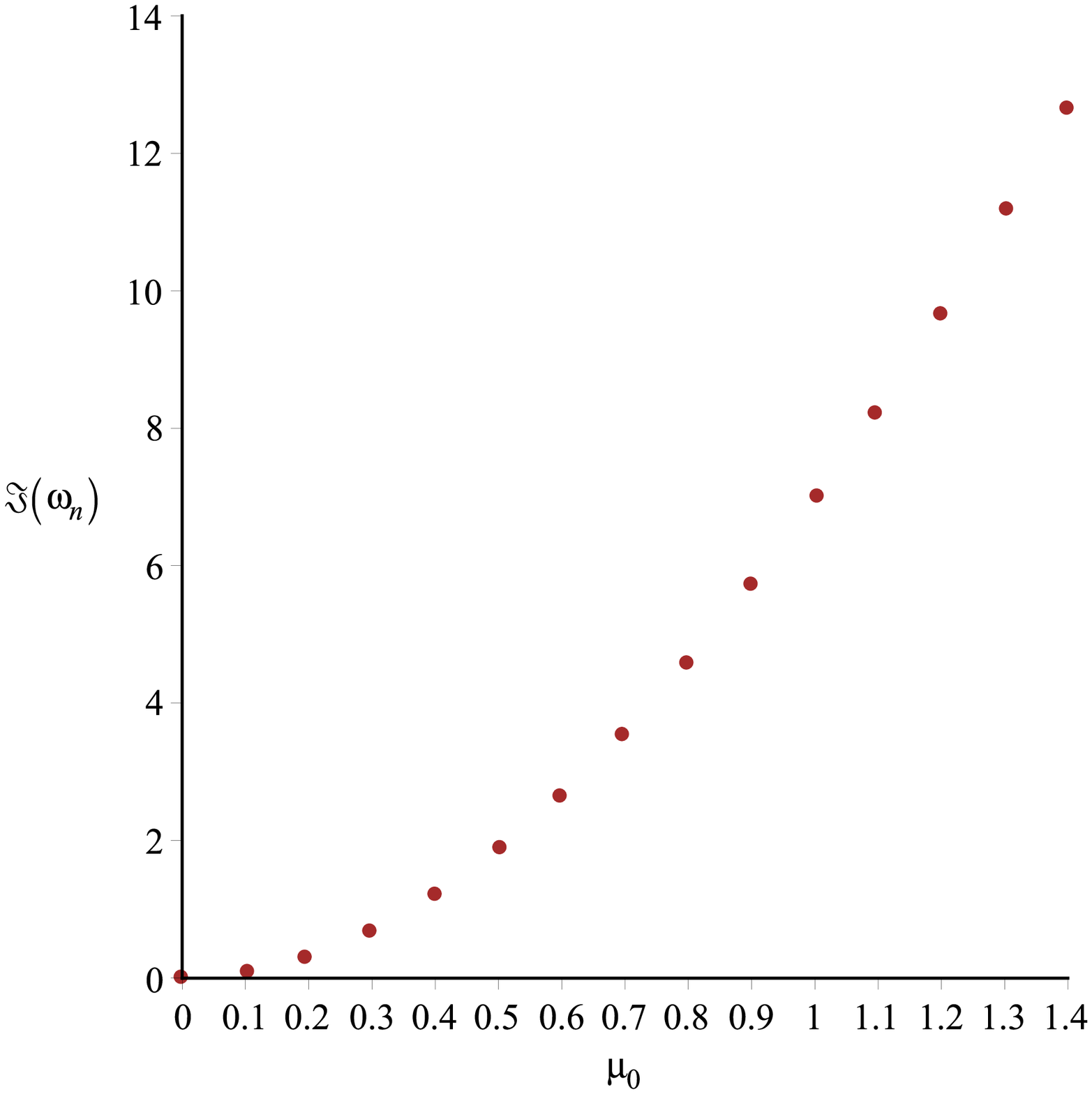}
	\figcaption{Scalar resonant frequencies for $n=1$, $l=0$, $\psi_{0}=0.02$, and $8\pi\eta^{2}=10^{-5}$. The units are in multiples of the total mass $M$.}
	\label{fig:Fig2_monopole}
%
		\includegraphics[scale=0.40]{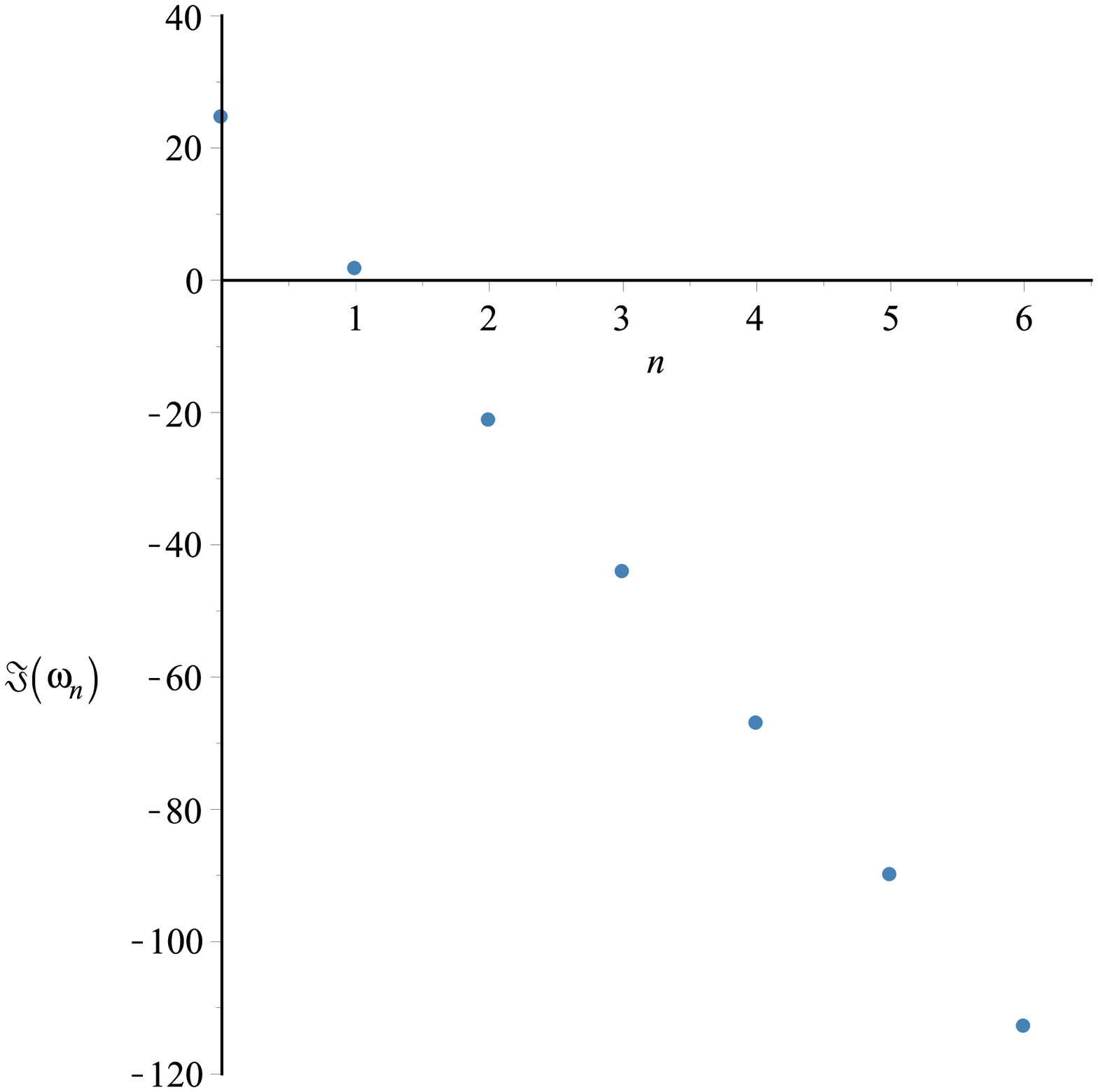}
	\figcaption{Scalar resonant frequencies for $l=1$, and $\mu_{0}=0.4$, $\psi_{0}=0.02$, and $8\pi\eta^{2}=10^{-5}$. The units are in multiples of the total mass $M$.}
	\label{fig:Fig3_monopole}
%
		\includegraphics[scale=0.40]{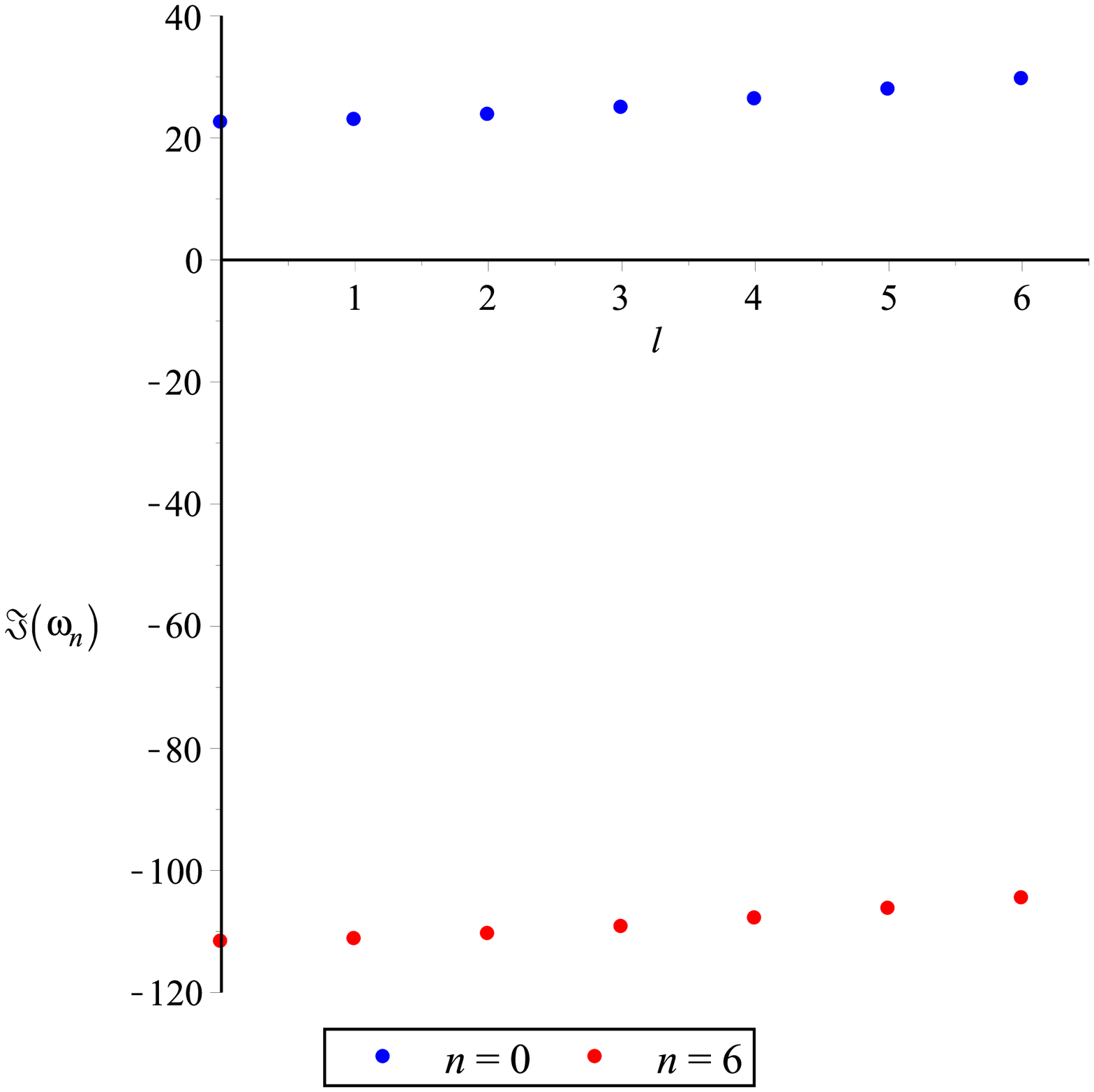}
	\figcaption{Scalar resonant frequencies for $\mu_{0}=0.1$, $\psi_{0}=0.02$, and $8\pi\eta^{2}=0.02$. The units are in multiples of the total mass $M$.}
	\label{fig:Fig4_monopole}
%
		\includegraphics[scale=0.40]{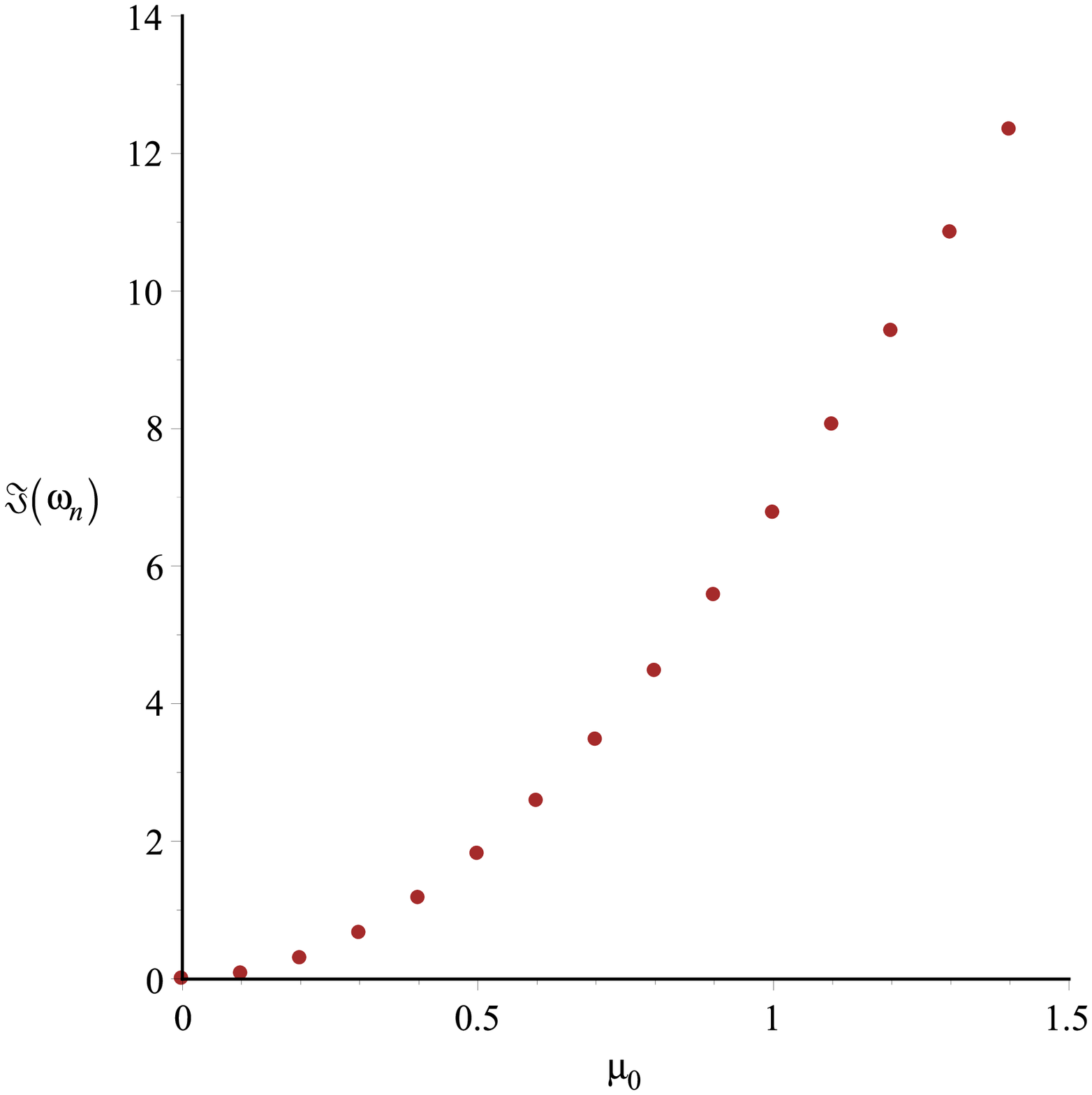}
	\figcaption{Scalar resonant frequencies for $n=1$, $l=0$, $\psi_{0}=0.02$, and $8\pi\eta^{2}=0.02$. The units are in multiples of the total mass $M$.}
	\label{fig:Fig5_monopole}
%
		\includegraphics[scale=0.40]{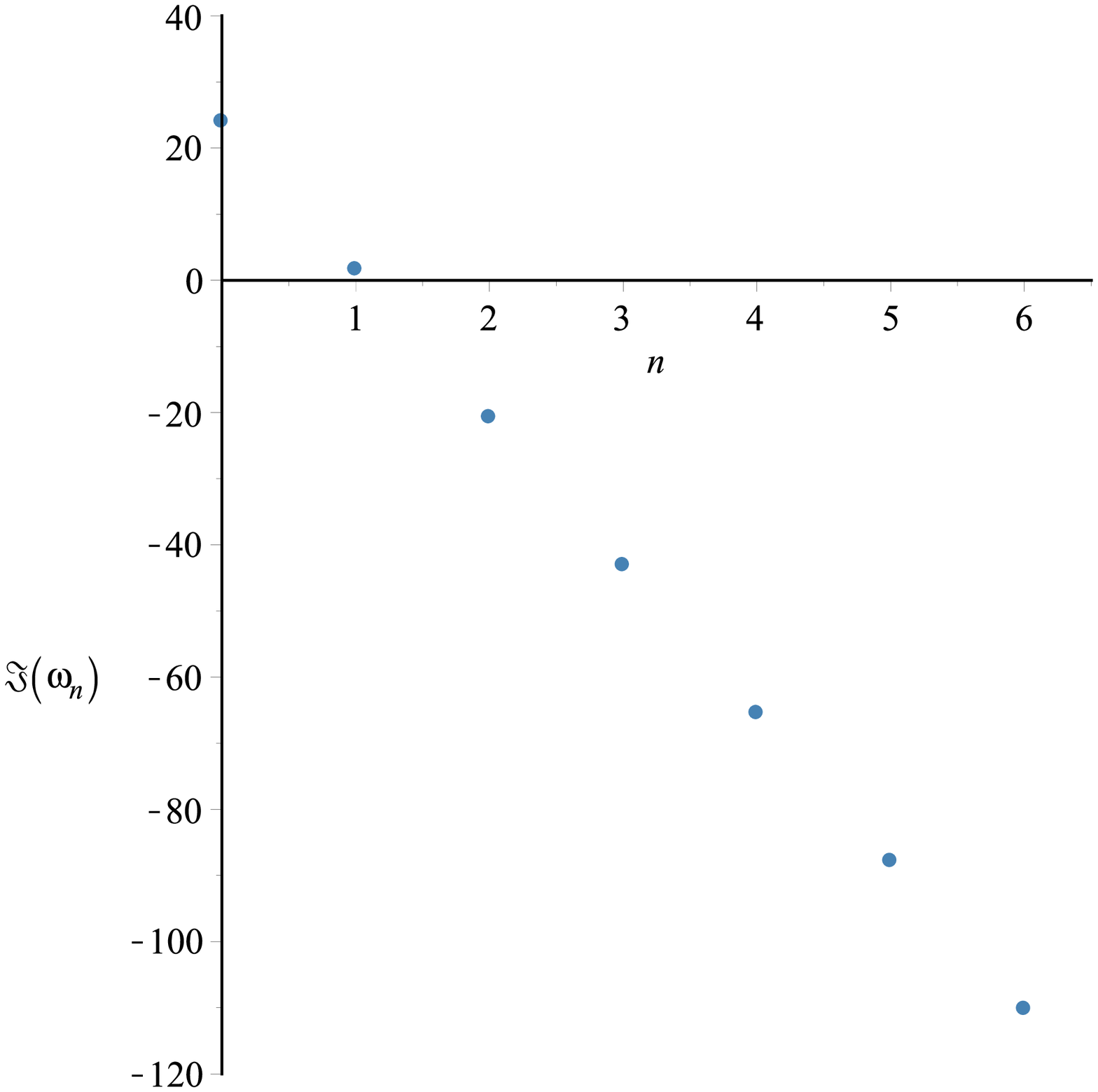}
	\figcaption{Scalar resonant frequencies for $l=1$, and $\mu_{0}=0.4$, $\psi_{0}=0.02$, and $8\pi\eta^{2}=0.02$. The units are in multiples of the total mass $M$.}
	\label{fig:Fig6_monopole}
\end{center}

Indeed, since the singularity $a$ is a constant given by Eq.~(\ref{eq:singularity_a_spherically_monopole}), the resonant frequencies given by Eq.~(\ref{eq:Energy_levels_spherically_monopole_x}) are not a function of $a$. However, for completeness, we will analyze the behavior of $\omega_{n}(a)$. This is shown in Figs.~(\ref{fig:Fig7_monopole}) and (\ref{fig:Fig8_monopole}).

\begin{center}
		\includegraphics[scale=0.40]{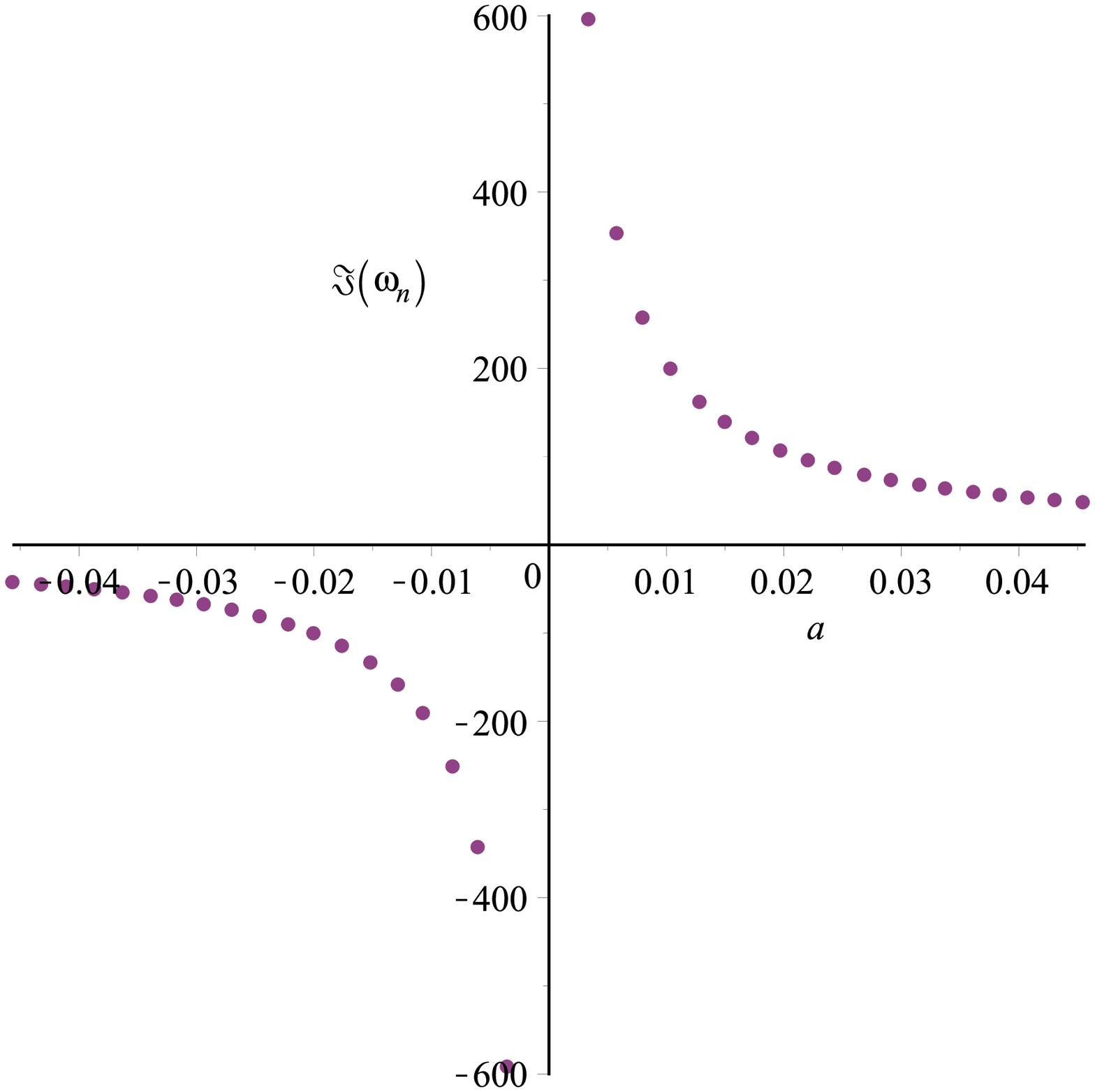}
	\figcaption{Values of the resonant frequencies for the mode $n=3$, $l=2$, $\mu_{0}=0.3$, $\psi_{0}=0.02$, $8\pi\eta^{2}=10^{-5}$ for different values of singularity $a$. The units are in multiples of the total mass $M$.}
	\label{fig:Fig7_monopole}
%
		\includegraphics[scale=0.40]{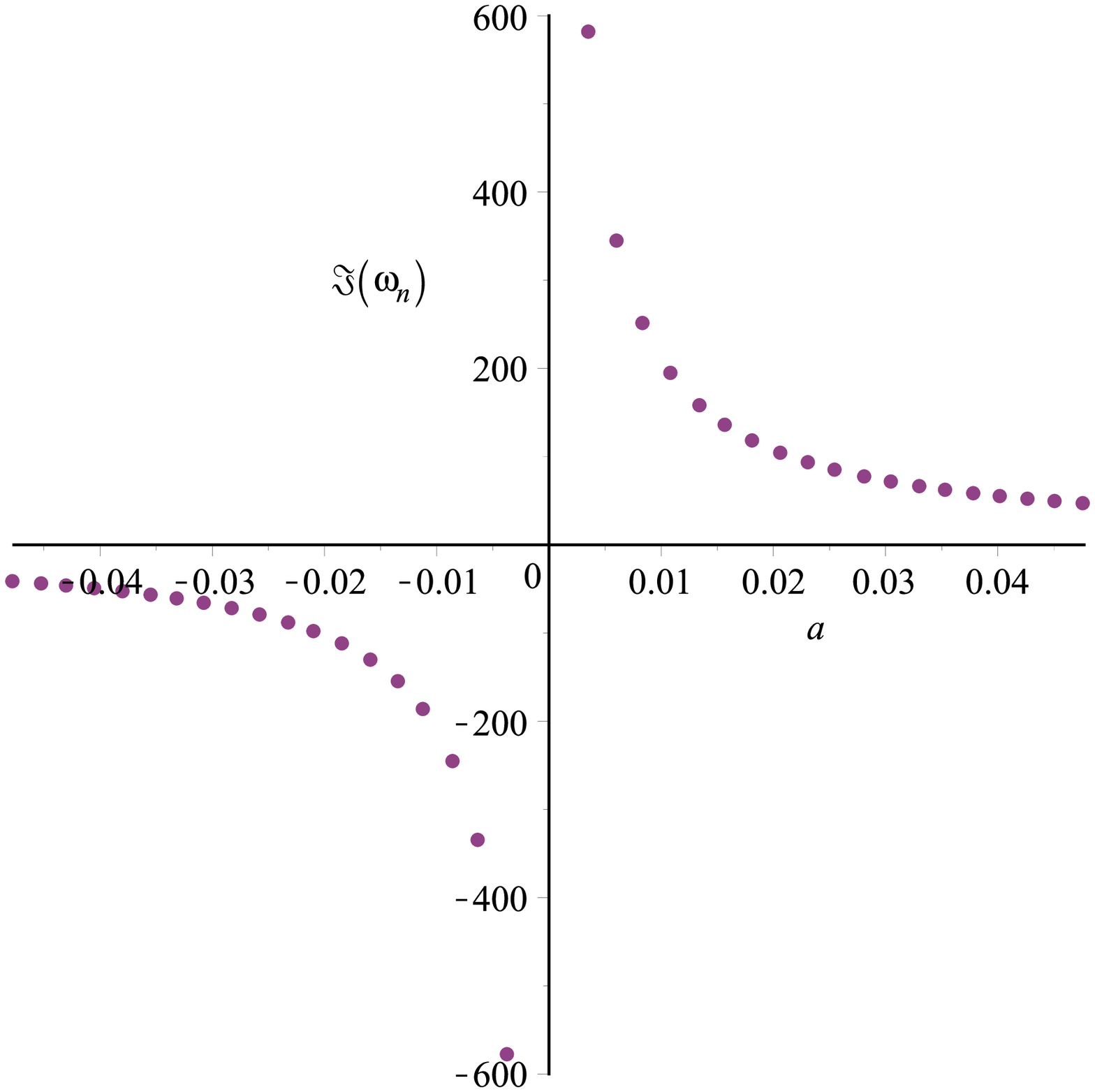}
	\figcaption{Values of the resonant frequencies for the mode $n=3$, $l=2$, $\mu_{0}=0.3$, $\psi_{0}=0.02$, $8\pi\eta^{2}=0.02$ for different values of singularity $a$. The units are in multiples of the total mass $M$.}
	\label{fig:Fig8_monopole}
\end{center}

From Figs.~(\ref{fig:Fig1_monopole})-(\ref{fig:Fig8_monopole}) we can conclude that there is no relevant difference between the values of the resonant frequencies concerning the two choices of the parameter $\eta$.
%
%
\section{Conclusions}
We have presented an exact solution for the radial part of the covariant Klein-Gordon equation for a massive scalar field in a Schwarzschild black hole with a global monopole in $f(R)$ gravity. We compare this result with those obtained in our recent works.

We used the general Heun functions to investigate some processes associated with scalar fields in the background under consideration, such as the existence of resonant frequencies and the Hawking radiation.

From this analytic solution, corresponding to the radial part, we obtained the solutions for ingoing and outgoing waves near the interior horizon, and used these results to discuss the Hawking radiation effect, by taking into account the properties of the general Heun functions. We also obtained a general expression for the energy flux for massive scalar particles.

Finally, we obtained the resonant frequencies. The RFs are a complex number and the imaginary component of the RFs tells us how quickly the oscillation will die away. Therefore, using these results, it is possible, in principle, to get some information about the physics of this black hole as well as to validate the $f(R)$ theory of gravity.

\vspace{15mm}

%
%


\vspace{-1mm}
\centerline{\rule{80mm}{0.1pt}}
\vspace{2mm}


%
%


\clearpage
\end{CJK*}
\end{document}